\documentclass[runningheads]{llncs}

\usepackage{tikz}
\usepackage{comment}
\usepackage{amsmath,amssymb} 
\usepackage{color}
\usepackage{color, colortbl}
\usepackage{adjustbox}
\usepackage{array, enumitem, soul, algorithm, algpseudocode}
\usepackage{graphicx}
\usepackage{cite}
\usepackage{xcolor}
\usepackage{xspace}
\usepackage{booktabs}
\usepackage{pifont}
\usepackage{url}
\usepackage{multirow}
\usepackage[pagebackref,breaklinks,colorlinks]{hyperref}

\definecolor{Gray}{gray}{0.92}
\newcommand{\xmark}{\text{\ding{55}}}%

\makeatletter
\DeclareRobustCommand\onedot{\futurelet\@let@token\@onedot}
\def\@onedot{\ifx\@let@token.\else.\null\fi\xspace}

\def\ie{\emph{i.e}\onedot}

\def\et{\emph{et al}\onedot}
\makeatother

\usepackage[accsupp]{axessibility}  

\begin{document}
\pagestyle{headings}
\mainmatter
\def\ECCVSubNumber{81}  

\title{Real-Time Under-Display Cameras\\Image Restoration and HDR on Mobile Devices} 

\titlerunning{Real-time UDC Image Restoration and HDR on Smartphones}
%
\author{Marcos~V.~Conde\inst{1} \and Florin Vasluianu\inst{1} \and Sabari Nathan\inst{2} \and Radu Timofte\inst{1}}
\authorrunning{Conde et al.}
%
\institute{
Computer Vision Lab, IFI CAIDAS, University of Würzburg, Germany
\email{\{marcos.conde-osorio,radu.timofte\}@uni-wuerzburg.de}\\
\url{https://github.com/mv-lab/AISP/}
\and
Couger Inc., Tokyo, Japan
}
\maketitle

\begin{abstract}
The new trend of full-screen devices implies positioning the camera behind the screen to bring a larger display-to-body ratio, enhance eye contact, and provide a notch-free viewing experience on smartphones, TV or tablets. On the other hand, the images captured by under-display cameras (UDCs) are degraded by the screen in front of them.
Deep learning methods for image restoration can significantly reduce the degradation of captured images, providing satisfying results for the human eyes. However, most proposed solutions are unreliable or efficient enough to be used in real-time on mobile devices.
In this paper, we aim to solve this image restoration problem using efficient deep learning methods capable of processing FHD images in real-time on commercial smartphones while providing high-quality results. We propose a lightweight model for blind UDC Image Restoration and HDR, and we also provide a benchmark comparing the performance and runtime of different methods on smartphones.
Our models are competitive on UDC benchmarks while using $\times4$ less operations than others.
To the best of our knowledge, we are the first work to approach and analyze this real-world single image restoration problem from the efficiency and production point of view.
\end{abstract}

\section{Introduction}
\label{sec:intro}

The consumer demand for smartphones with bezel-free, notch-less display has sparked a surge of interest from the phone manufacturers in a newly-defined imaging system, Under-Display Camera (UDC).
Besides smartphones, UDC also demonstrates its practical applicability in other scenarios, \ie, for videoconferencing with UDC TVs, laptops, or tablets, enabling more natural gaze focus as they place cameras at the center of the displays~\cite{msudc2020, zhou2020udc}. 
The UDC system consists of a camera module placed underneath and closely attached to the semi-transparent Organic Light-Emitting Diode (OLED) display~\cite{zhou2020udc}. This solution provides an advantage when it comes to the user experience analysis, with the full-screen design providing a higher level of comfort, and better display properties. The disadvantage of this solution is the fact that the OLED display acts as an obstacle for the light interacting with the camera sensor, inducing additional reflections, refractions and other connected effects to the Image Signal Processing (ISP)~\cite{conde2022modelbased} model characterizing the camera. 
Even though the display is partially transparent, there are opaque regions (i.e. the regions between the display pixels) that substantially affect the incoming light through a multitude of diffractions that are extremely difficult to model~\cite{msudc2020, zhou2020udc}. 

In this work, we aim to address the aforementioned issues. We present a novel solution for the UDC camera image restoration problem, that overcomes the difficulties induced by the real-world degradation in the UDC images. Our method provides high-quality results in real-time working conditions (under one second for image inference). We experiment with one of the world's first production UDC device, ZTE Axon 20, which incorporates a UDC system into its selfie camera~\cite{Feng_2021_CVPR_Discnet, feng2022mipi}. 

\noindent In summary, our contributions are as follows:
\begin{enumerate}
    \item We propose a new U-Net based architecture characterized by a lower number of parameters, enhancing its performance level by using an attention branch.
    \item We optimize our model in terms of used parameters, the number of FLOPS and the used operations, thus being able to achieve real-time performance on current smartphone GPUs at Full-HD input image resolution.
    \item We propose a new type of analysis, from a production point of view. This analysis looks at the behaviour of the different models when deployed on smartphone hardware platforms targeting different market segments and various price levels.  
\end{enumerate}


\section{Related Work}
\label{sec:rel-work}

In recent years, several works characterized and analyzed the diffraction effects of UDC systems~\cite{qin2017evaluation, tang202028, Feng_2021_CVPR_Discnet, msudc2020, Kwon_2021_CVPR, luo2022under}. Kwon \et \cite{kwon2016modeling} modeled the edge spread function of transparent OLED systems. Qin \et \cite{qin2016see} identified a pixel structure design that can potentially reduce the light diffraction. Additionally, several works \cite{suh2012p, suh201350} proposed UDC designs for enhanced interaction with flat displays.
Zhou \et{}~\cite{udc-ir-cvpr21} and their 2020 ECCV challenge \cite{zhou2020udc} were the first works that directly addressed this novel restoration problem using deep learning. Baidu Research team~\cite{zhou2020udc} proposed the Residual dense based on Shade-Correction for T-OLED UDC Image Restoration. They used the matching coefficient of the input image to do a light intensity correction, removing the input shade. 
In \cite{udc-ir-cvpr21}, the authors devised a Monitor Camera Imaging System (MCIS) to capture paired images, and solve the UDC image restoration problem as a blind deconvolution problem. Using the acquired data, they proposed a fully-supervised UNet~\cite{ronneberger2015u} like model for this task. More recently, Feng \et{}~\cite{Feng_2021_CVPR_Discnet} proposed one of the world's first production UDC device for data collection, experiments and evaluations. They also proposed a new model called DISCNet~\cite{Feng_2021_CVPR_Discnet} for non-blind restoration, and provided a benchmark for multiple blind and non-blind methods on their brand-new SYNTH dataset~\cite{Feng_2021_CVPR_Discnet}

However, the aforementioned UDC image restoration works~\cite{tang202028, Feng_2021_CVPR_Discnet, msudc2020, kwon2016modeling, qin2016see} suffer from several drawbacks.
The first drawback is introduced by the usage of an MCIS system to estimate the camera Point Spread Function (PSF)~\cite{joshi2008psf}, given the acquisition of non-realistic paired data to develop a model \cite{peng2019learned, asif2016flatcam}. In ~\cite{liu2020singlehdr}, the authors identified MCISs limitations in terms of High Dynamic Range (HDR), which is crucial for a realistic diffraction artifacts removal. 
The second drawback comes as the authors are trying to simulate a UDC system by manually covering a camera with an OLED display. Even if this setup provides them with quasi-realistic data, the properties of an actual UDC system are much more complex than this approximation. Regular events like tilting, regular rotations or camera motion will produce a PSF that varies during recording, and thus, their models will suffer when asked to handle different degradations (as they will rely on a rough approximation of the PSF given by the simulated data). In ~\cite{Feng_2021_CVPR_Discnet}, Feng \et{} proposed the usage of one of the first commercially available UDC devices for data collection, enabling realistic data conducted experiments, that further enhanced the performance of the evaluated UDC image restoration models.

Lately, the tendency has switched to focus the efforts in computing reliable PSFs, to be then used in data synthesis. This enables solving the UDC image restoration in the fully supervised framework, with various UNet-like models being used as image-to-image translators. However, the real world domain PSF is still able to provide useful information for PSF based models. For example,  Feng \et{}~\cite{Feng_2021_CVPR_Discnet} proposed DISCNet, a method which achieved state-of-the-art results by leveraging the computed PSF to improve performance.  They provide a benchmark for multiple blind and non-blind methods, developed using their real synthetic UDC data ~\cite{Feng_2021_CVPR_Discnet}.

\paragraph{\textbf{Non-blind Image Restoration.}}
The Non-Blind Image Restoration implies the usage of the real PSF characterizing the system in the model design. Once the PSF is known, the deconvolution approach can be used. Several works have follow this approach \cite{orieux2010bayesian-wiener,levin2009understanding,cho2011handling,whyte2014deblurring}, where the authors imposed prior knowledge regarding the acquisition system to limit the solution space size for the estimation of the unknown noise model. Lately, different authors  \cite{schuler2013machine,xu2014deep,zhang2017learning} have focused their efforts in the design of neural networks aiming for non-blind image restoration. For example, Zhang \et \cite{zhang2018learning} proposed SRMD, using a single neural network to handle multiple degradations. Gu \et \cite{gu2019blind} proposed a method called SFTMD, using the Iterative Kernel Correction (IKC) to gradually correct the existing degradations. Generative Adversarial Networks (GANs) were used in various works \cite{bell2019blind,yuan2018unsupervised,zhou2019kernel}, to produce a more realistic reconstruction by tackling the remaining unknown degradations which are not described by the PSF.  
The possibility of using the PSF kernel in the model design was explored in works like \cite{zhang2018learning}. To date, Feng \et DISCNet~\cite{Feng_2021_CVPR_Discnet} is the state-of-the-art for non-blind UDC image restoration. This method defines a physics-based image formation model to better understand the degradation.  The authors measure the real-world Point Spread Function (PSF) from smartphone prototypes, and design a Dynamic Skip Connection Network (DISCNet) to restore the UDC images. This approach shows strong results on both synthetic and real data samples.

\paragraph{\textbf{Blind Image Restoration.}} 
Solutions tackling the Blind Image Restoration problem do not rely on any prior knowledge about the Point Spread Function (PSF) to describe the light transport effects characteristic to a UDC setup. This makes impossible the usage of the PSF for data synthesis, and the amount of available training data becomes another limitation of this difficult problem. This type of solution was proposed by Zhou \et{}~\cite{udc-ir-cvpr21, zhou2020udc} (\ie{} DE-UNet). Most recently Koh \et{} \cite{Koh_2022_CVPR} use a two-branch neural network tackling specifically the light diffraction produced by the pixels grid, and the diffuse intensity caused by the additional film layers used for the display system. 

\vspace{2mm}

\noindent However, none of all the previous mentioned methods have analyzed the problem from an \textbf{efficiency} point of view. We believe this is due to the fact that the current proposed models can generate high-quality results but cannot be integrated into modern smartphones due to their complexity (\ie{ number of FLOPs, memory requirements}).


\section{Proposed Method}

We propose two solutions for tackling the UDC image restoration problem.  Both solutions are blind image restoration methods, and therefore, they do not rely on the PSF information about the camera:

\begin{enumerate}
    \item \textbf{DRM-UDCNet}: A UNet-like model inspired in previous solutions~\cite{Feng_2021_CVPR_Discnet, udc-ir-cvpr21, msudc2020, zhou2020udc}. For this network, we propose a 2.2 Million parameters dual-branch model that employs our own formulation of Residual Dense Attention Blocks (RDAB)~\cite{zhang2018residualdense}. This represents a reduction of 40\% of the number of parameters with respect to the models that achieved the current state-of-the-art results \cite{Feng_2021_CVPR_Discnet, udc-ir-cvpr21}.
    \vspace{2mm}
    \item \textbf{LUDCNet}: A compact version of \emph{DRM-UDCNet}, tailored specifically to be efficient when deployed on various smartphone devices. As this model is designed to perform UDC image restoration in real-time conditions on mobile devices, we optimize it considering SSIM performance and a lower number of FLOPs. As we will prove in Section~\ref{sec:eff}, this model achieves competitive performance while processing Full-HD images in real-time on a wide range of commercial smartphones.
\end{enumerate}

Our methods combine ideas from \textbf{deblurring}~\cite{nah2021ntire} and \textbf{HDR}~\cite{liu2020singlehdr, catley2022flexhdr} networks, and attention methods~\cite{woo2018cbam, liu2018intriguingcoord}. We believe these tasks are extremely correlated with the UDC restoration problem and applications. Our method is illustrated in Figure~\ref{fig:main}. The designed dual-branch model allows to adapt it depending on the trade-off between resources and performance. For instance, we can remove the attention branch and obtain competitive results with a lower inference time. In exchange of slightly higher inference time, we can improve notably our performance by using the proposed attention branch.

\subsection{Method Description}

\begin{figure}
    \centering
    \includegraphics[width=\linewidth]{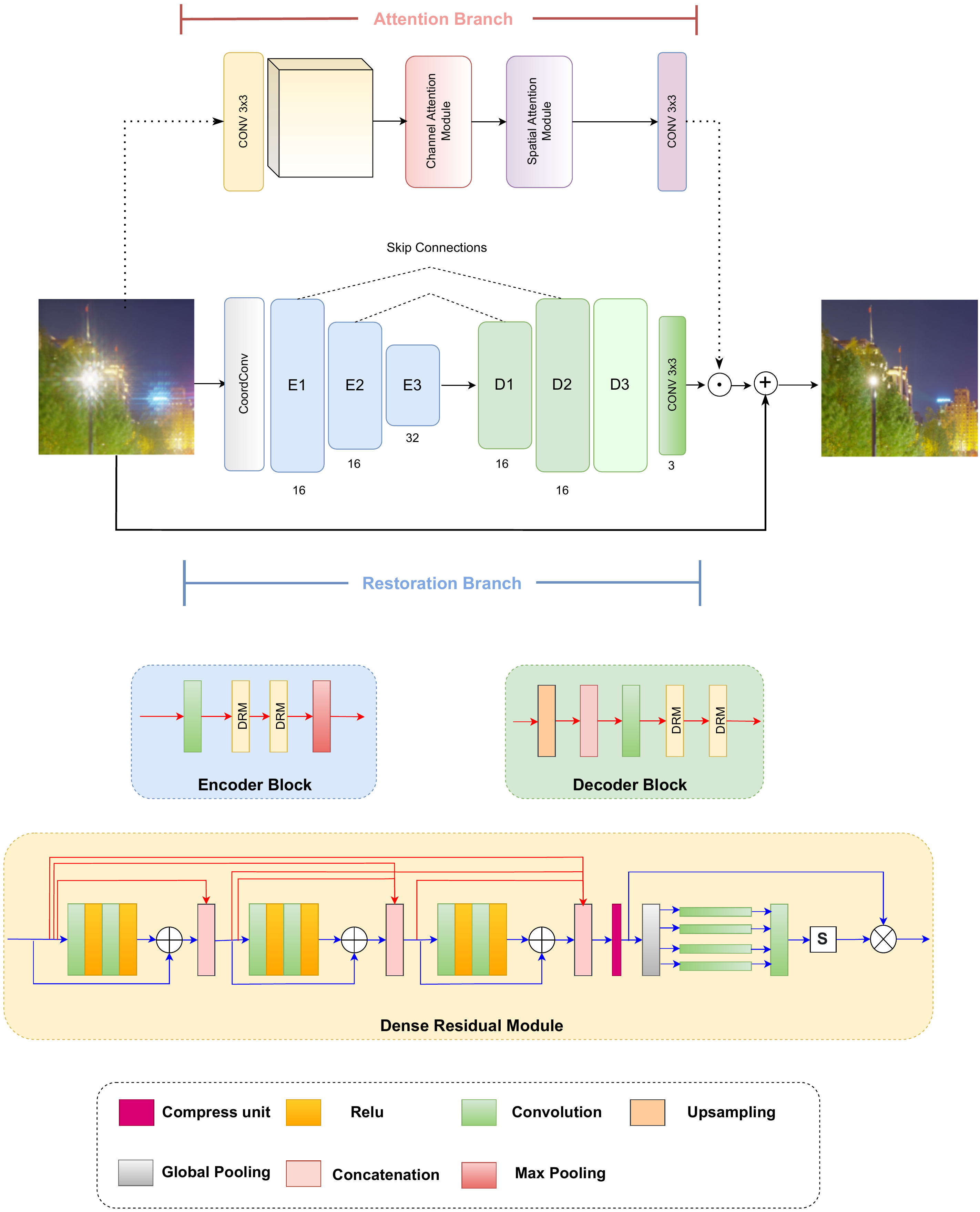}
    \caption{Architecture of the proposed network. The top of the figure shows the overall \emph{DRM-UDCNet} model for blind UDC Image Restoration. The bottom of the figure presents the Dense Residual Module~\cite{zhang2018residualdense}. Note that we apply pruning and sequentially remove layers to reduce its size after training. We provide more details in the supplementary material.}
    \label{fig:main}
\end{figure}

As we show in Figure~\ref{fig:main}, \emph{DRM-UDCNet} consists of:

\begin{enumerate}
    \item The main image restoration branch: It is composed of three encoder blocks ($E_1$, $E_2$, $E_3$) and three decoder blocks ($D_1$, $D_2$, $D_3$). Each encoder block consists of two Dense Residual Modules (DRM)~\cite{zhang2018residualdense,ronneberger2015unet} and a downsampling layer. 
    The decoder blocks $D_1$, $D_2$ have two DRM followed by a bilinear upsampling layer.
    Note that the input image spatial coordinates are mapped using the Coordinate Convolution Layer~\cite{liu2018intriguingcoord}, and we use skip connections between encoder-decoder blocks~\cite{ronneberger2015unet}. The output of the final decoder block is mapped into a 3-channel residual using a convolution layer and \emph{tanh} activation. We further activate this residual, multiplying it by an attention map $\mathcal{A}$. This is a technique used in state-of-the-art single image HDR~\cite{liu2020singlehdr}. We finally apply global residual learning as a standard technique in image restoration. 
    
    \item Our attention branch: Similar to HDR methods~\cite{liu2020singlehdr}, it aims at generating a 3-channel attention map to control artifacts and hallucination in overexposed areas. The attention map is generated after applying a CBAM block~\cite{woo2018cbam} to the input image. The final layer is activated using a sigmoid function to produce a per-channel attention map $\mathcal{A}$.
    
\end{enumerate}

We further compact our challenge model \emph{DRM-UDCNet} into \emph{LUDCNet}, which only uses two DRM blocks and a reduced number of filters. We remove Batch Normalization (BN) layers, which consume the same amount of GPU memory as convolutional layers, and they also increase computational complexity~\cite{zhang2018residualdense}. Furthermore, we replace classical CNN blocks by Inverted Linear Residual Blocks~\cite{howard2017mobilenets}. \emph{LUDCNet} will process the images at half-resolution, with the input image being downsampled by a factor of two, and the output image upsampled to the original resolution.

\section{Experimental Setup}

\subsection{Implementation details}

The original RGB images from SYNTH~\cite{Feng_2021_CVPR_Discnet} have resolution $800\times800$, we extracted non-overlapping patches of size $400\times400$. We extracted patches (instead of downsampling the image) to preserve the high-frequency details. Moreover, images were transformed from the original domain to the tone-mapped domain using the following function $f(x) = x / (x+0.25)$. Therefore the pixel intensity is in the range $[0,1)$. To avoid artifacts, the output of our models is clipped into range $[0,1-p]$ , where $p$ is $10^{-5}$. Our models were implemented in Tensorflow 2 and trained using a single Tesla P100 GPU (16Gb). Some experiments were performed using a TPU v3-8.
We used Adam optimizer with default hyper-parameters, and an initial learning rate of 1e-3. The learning rate is reduced by 50\% during plateaus up to a minimum learning rate of 1e-6. 
We use basic augmentations: horizontal flip, vertical flip, and rotations. We set 4 as mini-batch size and trained using only the SYNTH dataset~\cite{Feng_2021_CVPR_Discnet} to convergence for a few days.

\paragraph{Loss function}

The model is trained using a weighted sum of $\mathcal{L}1$ loss, SSIM loss, and gradient loss. We define the loss functions as follows: 
\begin{equation}
\label{eqn:4}
    \mathcal{L}_1 = \frac{1}{N}\sum_{i=1}^{N}\left \| \mathcal{I} - \mathcal{\hat{I}} \right \|_1
\end{equation}
where $\mathcal{D}$ is the input degraded image, $f(\mathcal{D}) = \hat{\mathcal{I}}$ is the restored image using our model $f$, and $\mathcal{I}$ is the ground truth image. The SSIM loss is defined as:

\begin{equation}
\label{eqn:5}
      \mathcal{L}_{SSIM} = \frac{1}{N}\sum_{i=1}^{N}(1-SSIM(\mathcal{I}, \hat{\mathcal{I}}))
\end{equation}
where SSIM is the structural similarity index function defined in~\cite{SSIMPaper}.

The gradient loss is the $\mathcal{L}_1$ distance between the gradients of the restored image and the ground-truth, in the $xy$ axis:
\begin{equation}
\label{eqn:6}
 \mathcal{L}_{Grad} = \frac{1}{N}\sum_{i=1}^{N} \left \|\nabla_{x}\mathcal{I} - \nabla_{x}\hat{\mathcal{I}} \right \|_1 + \left \| \nabla_{y}\mathcal{I} - \nabla_{y}\hat{\mathcal{I}} \right \|_1
 \end{equation}
Therefore, the final loss function is:
\begin{equation}
\label{eqn:7}
    \mathcal{L} = 0.1~\mathcal{L}_{SSIM}+\mathcal{L}_1+\mathcal{L}_{Grad}
\end{equation}

\subsection{Experimental Results}

\subsubsection{Quantitative Results}

We evaluate our models on two different benchmarks: (i) the SYNTH dataset~\cite{Feng_2021_CVPR_Discnet}, also used in the ``UDC MIPI 2022 Challenge"~\cite{feng2022mipi}; (ii) the T-OLED UDC 2020 Dataset~\cite{msudc2020, zhou2020udc}. We will mainly focus on the SYNTH dataset~\cite{Feng_2021_CVPR_Discnet}, where we compare against current state-of-the-art methods (excluding contemporary methods, and methods better than DISCNet~\cite{Feng_2021_CVPR_Discnet} in the 2022 Challenge).
Note that these results are not fully reproducible since multiple approaches do not provide open-sourced code. Moreover, the SYNTH dataset~\cite{Feng_2021_CVPR_Discnet} test set (and PSF) is public, which makes difficult a fair comparison.
As we show in Tables~\ref{tab:results} and~\ref{tab:toled}, we achieve competitive results in both benchmarks, while using models with fewer parameters than the other competitive methods. As an example, we can look at the comparison against  DISCNet~\cite{Feng_2021_CVPR_Discnet} base version, as we used the same initial setup to build up our model. We can see how our method outperforms this one by almost 2dBs. We extend this analysis in Section~\ref{sec:eff}.

\begin{table}[!ht]
    \centering
    \caption{Quantitative Results for Under-display Camera (UDC) Image Restoration using the SYNTH dataset~\cite{Feng_2021_CVPR_Discnet}. We show fidelity and perceptual metrics. (*) indicates methods proposed at the 2022 UDC Challenge~\cite{feng2022mipi}. We consult some numbers from~\cite{Feng_2021_CVPR_Discnet}.}
    \label{tab:results}
    \resizebox{\linewidth}{!}{
    \begin{tabular}{l|c|c|c|c|c}
         \hline
         Method & PSNR~(dB)~$\uparrow$ & SSIM~$\uparrow$ & LPIPS~$\downarrow$  &\# Params & Uses PSF\\
         \hline
         Wiener Filter (WF)~\cite{orieux2010bayesian-wiener} & 27.30 & 0.83 & 0.330 & - & yes \\
         SRMDNF~\cite{Feng_2021_CVPR_Discnet} & 34.80 & 0.96 & 0.036 & 1.5 & yes \\
         SFTMD~\cite{Feng_2021_CVPR_Discnet} & 42.35 & 0.98 & 0.012 & 3.9 & yes \\
         \textbf{DISCNet (PSF)}~\cite{Feng_2021_CVPR_Discnet} & 42.77 & 0.98 & 0.012 & 3.8 & yes \\
         \hline
         RDUNet~\cite{zhang2018residualdense, ronneberger2015unet}& 34.37 & 0.95 & 0.040 & 8.1 & no\\
         DE-UNet~\cite{udc-ir-cvpr21} & 38.11 & 0.97 & 0.021 & 9.0 & no \\
         \textbf{DISCNet (w/o PSF)}~\cite{Feng_2021_CVPR_Discnet} & 38.55 & 0.97 & 0.030 & 2.0 & no \\
         RushRushRush~* & 39.52 & 0.98 & 0.021 & - & no\\
         eye3~* & 36.69 & 0.97 & 0.032 & - & no\\
         FMS Lab~* & 35.77 & 0.97 & 0.045 & - & no\\
         EDLC2004~* & 35.50 & 0.96 & 0.045 & - & no\\
         SAU\_LCFC~* & 32.75 & 0.96 & 0.056 & - & no\\
         DRM-UDCNet & 37.45 & 0.97 & 0.030 & 2.0 & no\\
         \rowcolor{Gray} \textbf{DRM-UDCNet + Attn} & 40.21 & 0.98 & 0.020 & 2.9 & no\\
         \hline
    \end{tabular}}
\end{table}

\begin{table}[!h]
    \centering
    \caption{Quantitative Results and comparison of methods at the T-OLED UDC 2020 Challenge~\cite{msudc2020, zhou2020udc}. \textbf{TT} denotes the training time, and \textbf{IT} the inference time per image. Note that we only compare with simple methods trained in similar conditions.}
    \label{tab:toled}
    \resizebox{\textwidth}{!}{
    \begin{tabular}
    {l|c|c|c|c|c}
    \hline
    {Team} & PSNR~(dB)~$\uparrow$ & SSIM~$\uparrow$ & TT(h) & IT (s/frame)  & {CPU/GPU} \\
    \hline
    CILab IITM  &$36.91$	&$0.9734$& 96  &1.72  & 1080 Ti  \\
    lyl &$36.72$	&$0.9776$&72  & 3.0  & -  \\
    \rowcolor{Gray} \textbf{DRM-UDCNet} &$36.50$	&$0.9730$ &14  &1.0  & Tesla P100  \\
    Image Lab &	$34.35$	&$0.9645$&-  &1.6  & 1080 Ti \\
    San Jose Earthquakes &$33.78$	&$0.9324$& 18  &180  &- \\
    UNet~\cite{zhou2020udc} & $32.42$ & $0.9343$ & 24  & -  & Titan X \\
    DeP~\cite{zhou2020udc} &$28.50$	&$0.9117$& 24  & -  & Titan X \\
    \hline
    \end{tabular}} 
\end{table}

\subsubsection{Qualitative Results}
\begin{figure}[!ht]
    \centering
    \setlength{\tabcolsep}{2.0pt}
    \begin{tabular}{c}
    \includegraphics[width=\linewidth]{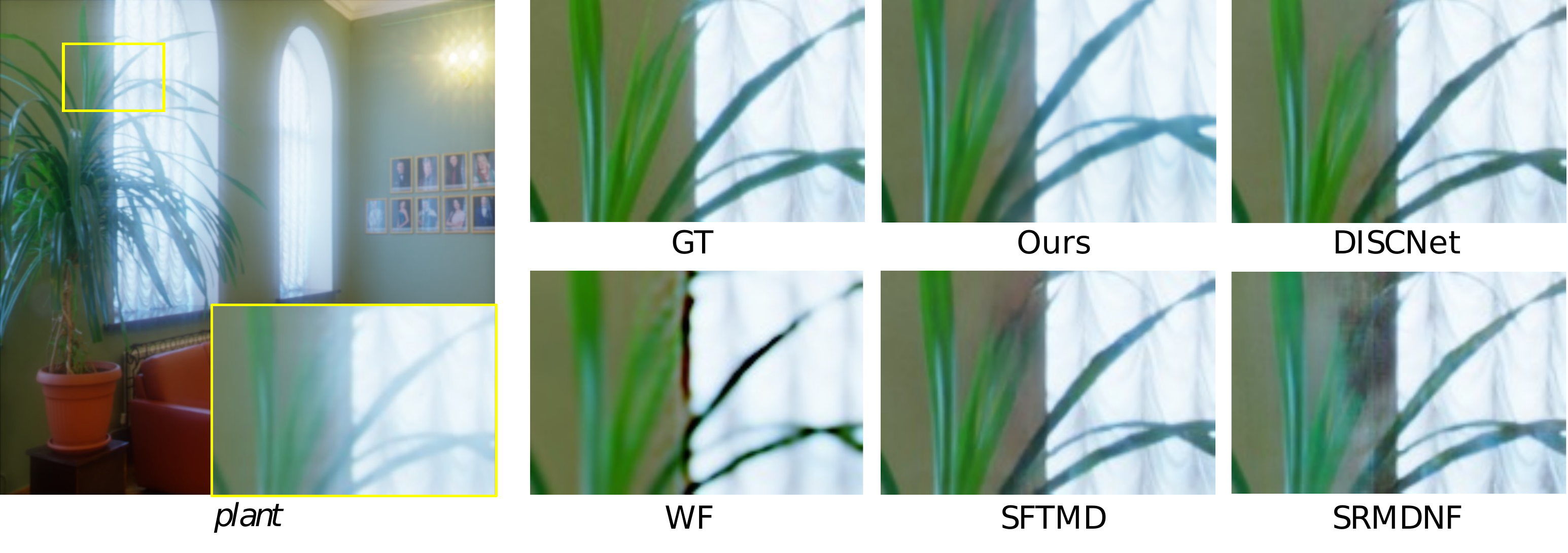} \tabularnewline
    \includegraphics[width=\linewidth]{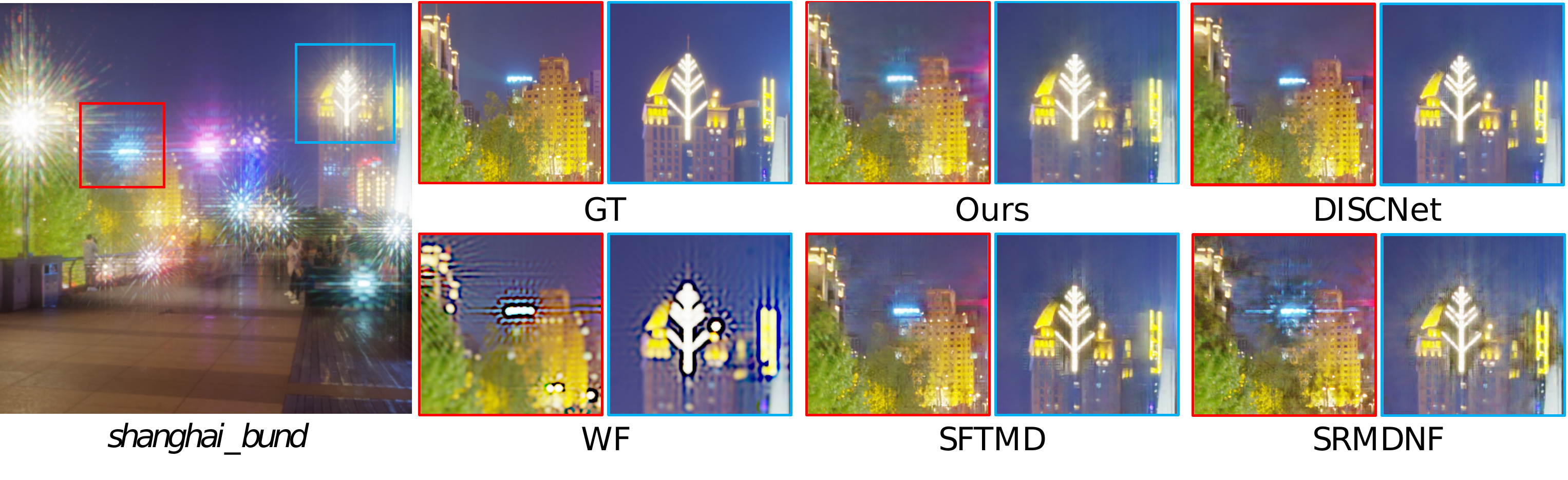}\tabularnewline
    \end{tabular}
    \caption{Visual comparison on synthetic validation images. Our method restores fine details, recovers information from HDR, and renders high perceptual quality results without notable artifacts. Images selection from DISCNet~\cite{Feng_2021_CVPR_Discnet}. Zoom in for better view.}
    \label{fig:dicnet-comp}
\end{figure}

\begin{figure}[!ht]
    \centering
    \setlength{\tabcolsep}{2.0pt}
    \begin{tabular}{cccc}
    \includegraphics[width=0.24\linewidth]{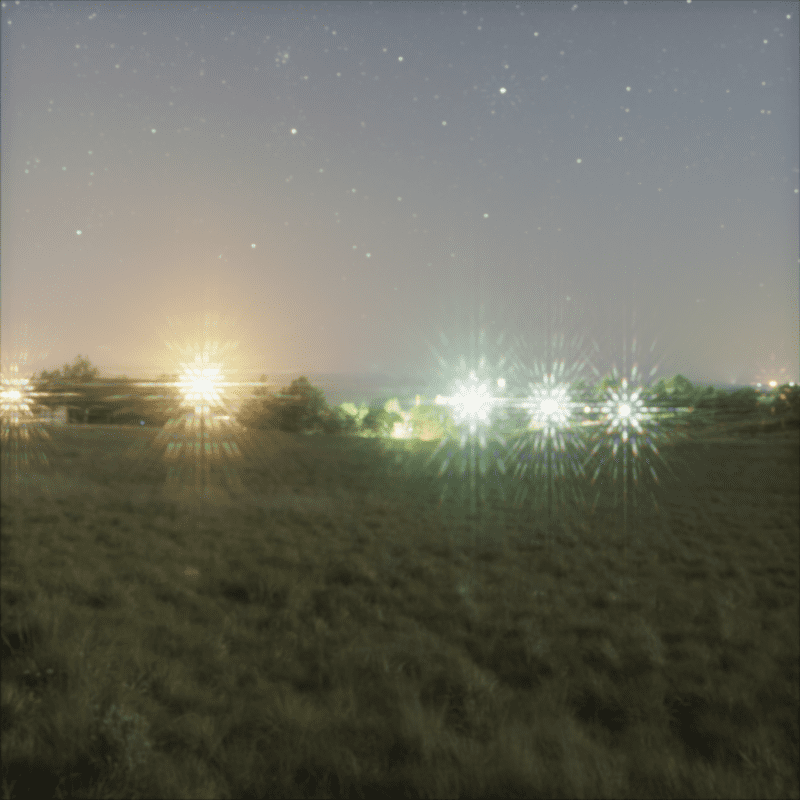} &
    \includegraphics[width=0.24\linewidth]{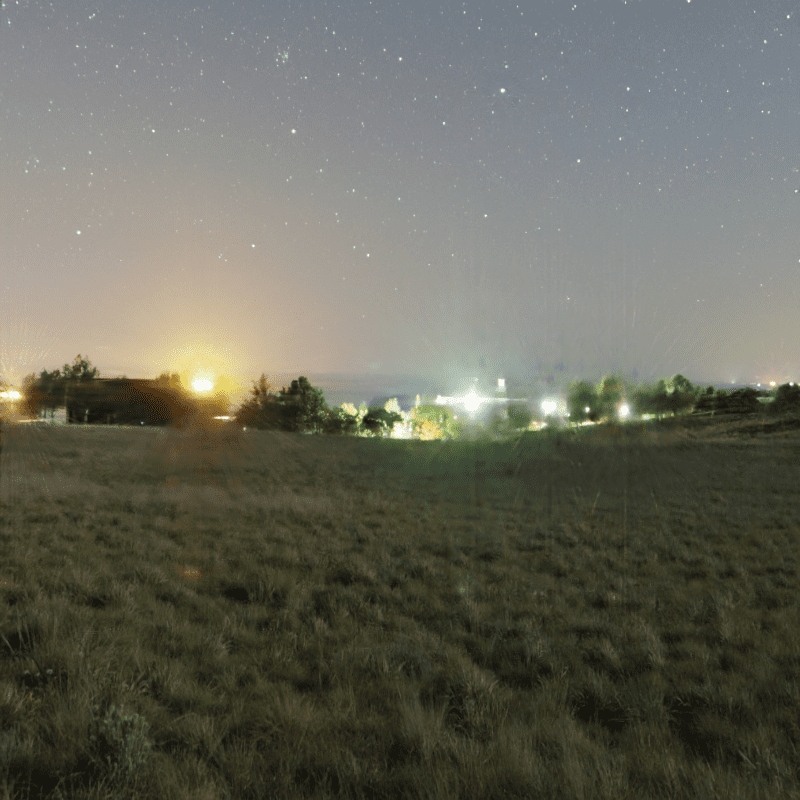} &
    \includegraphics[width=0.24\linewidth]{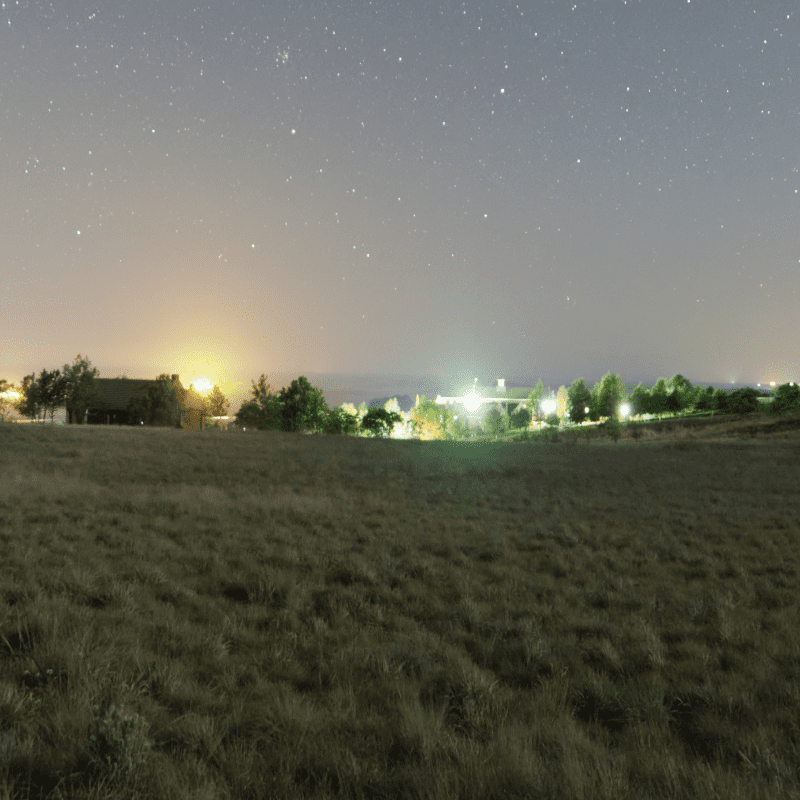} &
    \includegraphics[width=0.24\linewidth]{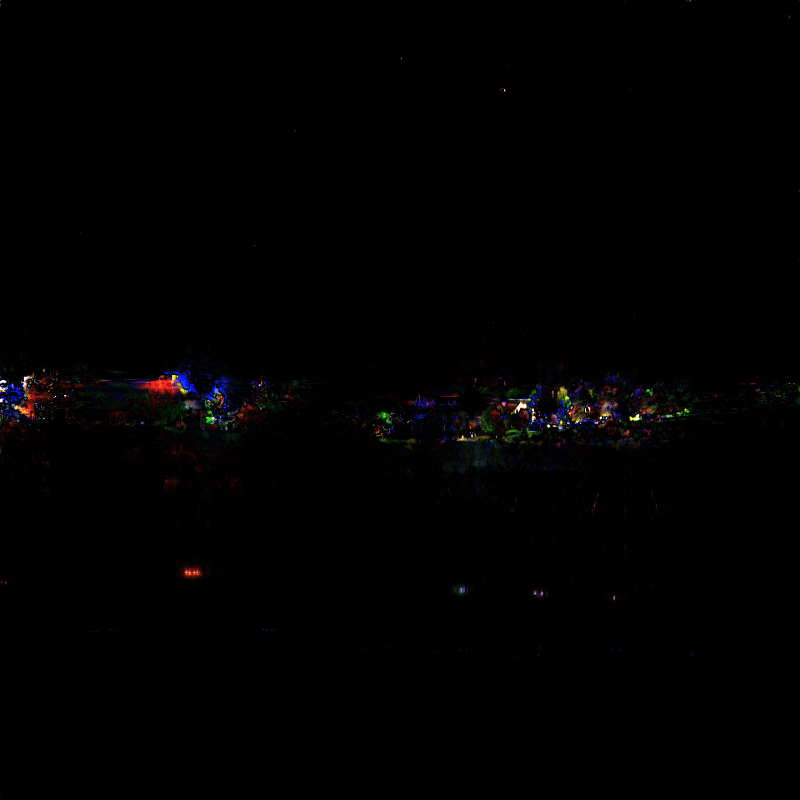} \tabularnewline
    \includegraphics[width=0.24\linewidth]{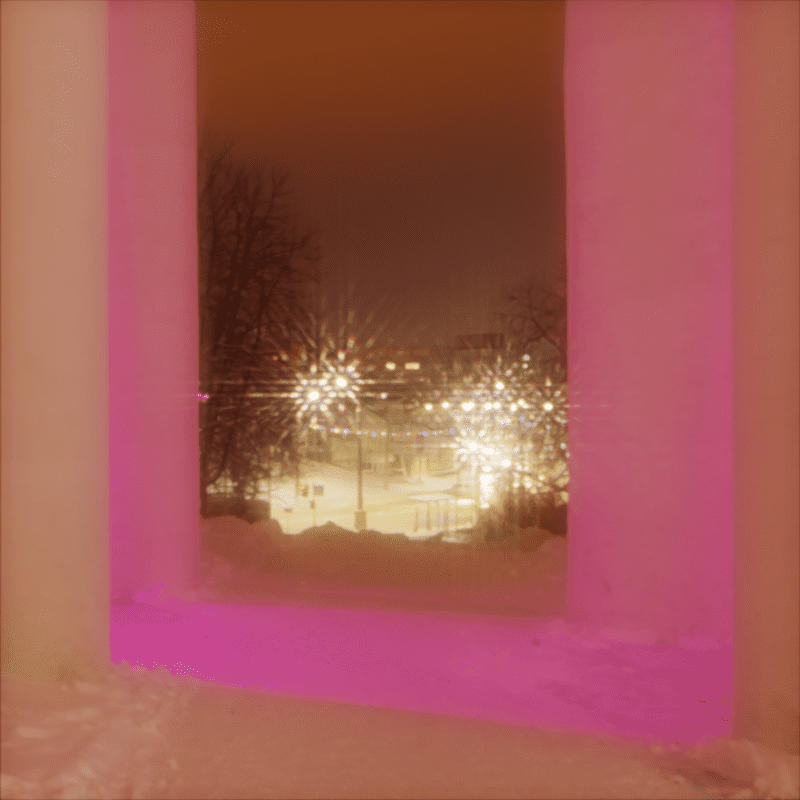} &
    \includegraphics[width=0.24\linewidth]{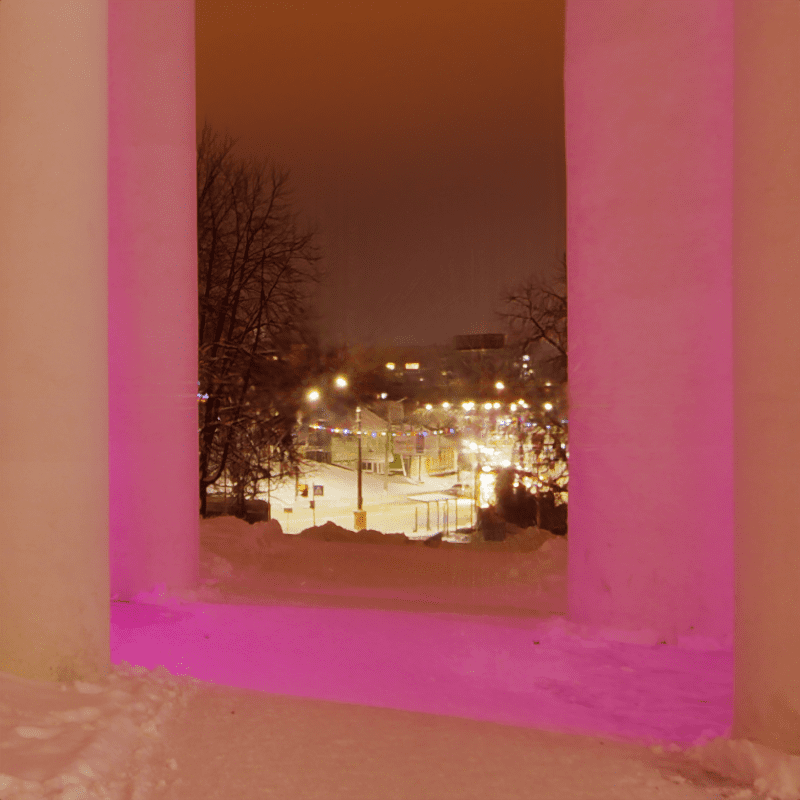} &
    \includegraphics[width=0.24\linewidth]{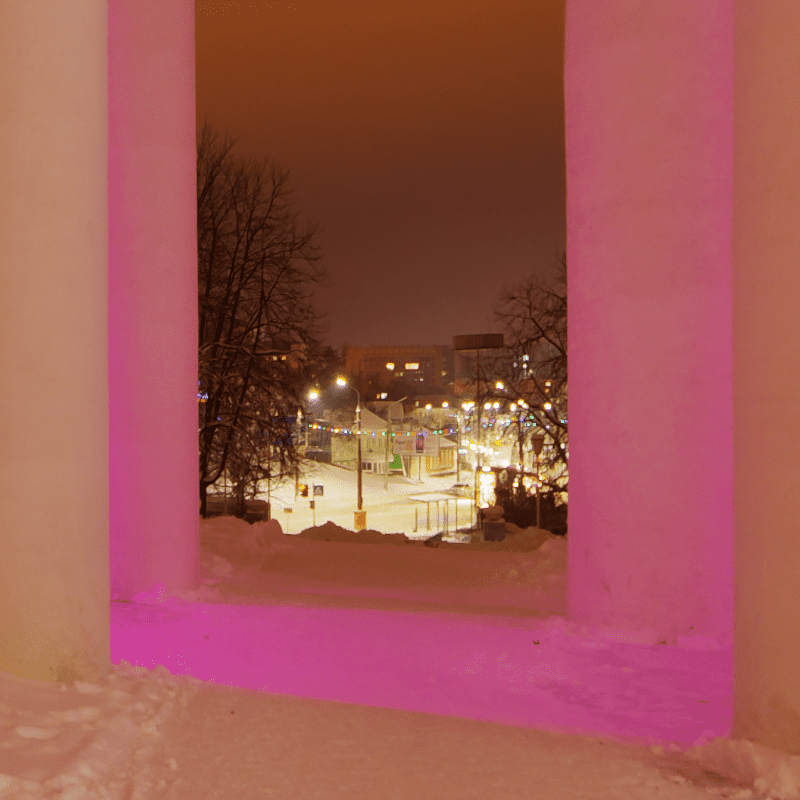} &
    \includegraphics[width=0.24\linewidth]{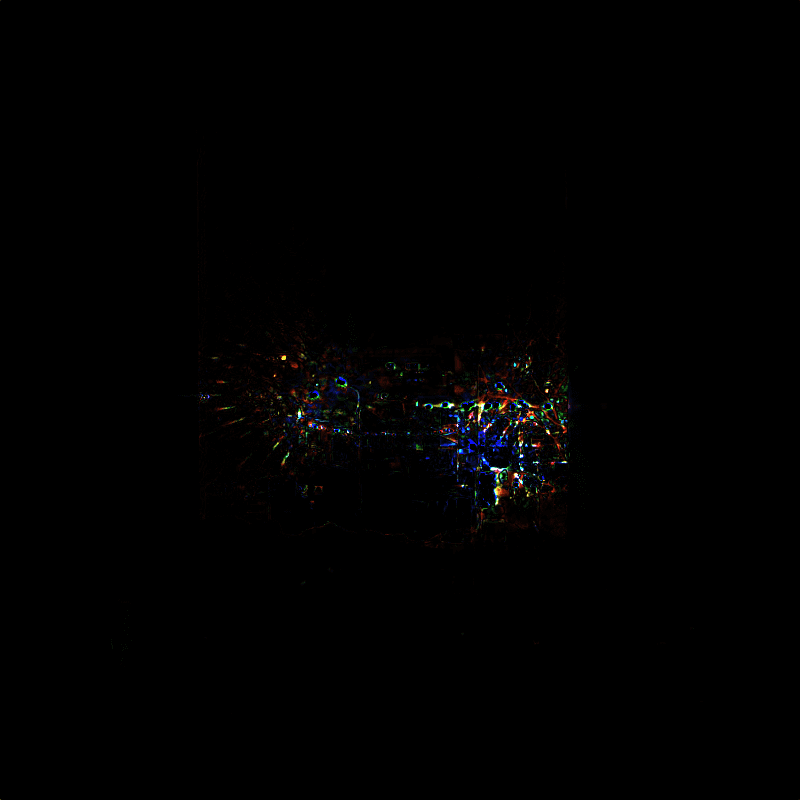} \tabularnewline
    Input & Prediction & Reference & Error Map \tabularnewline
    \end{tabular}
    \caption{SYNTH Dataset~\cite{Feng_2021_CVPR_Discnet} validation samples. Images visualized after applying tone-mapping. The error maps are multiplied by $16$ for a better visualization.}
    \label{fig:val-samples}
\end{figure}

Figure \ref{fig:dicnet-comp} provides a visual comparison for the results of our model, compared to other state-of-the-art methods. As you can see, our method is able to provide a better reconstruction, with results characterized by better properties in terms of textures, colors and the geometric properties of the hallucinated objects. In Figure \ref{fig:val-samples}, we provide equivalent images for a comparison of the model prediction to the ground truth images, given each input sample. On the last column (Fig. \ref{fig:val-samples}), we provide scaled error maps computed between the model prediction and the ground truth, observing that the most of the reconstruction error is concentrated in areas characterized by high light intensity, and therefore affected by over-exposure and glare. 

\section{Efficiency Analysis}
\label{sec:eff}

We evaluate the performance of different methods using three smartphone devices.
In Table~\ref{tab:phones}, we provide all the details regarding the target devices used in our benchmark. We choose two different smartphones that target the \emph{entry-level} segment of the  market (1,2), and we also consider a mobile phone designed for the \emph{flagship} segment (3). To observe the evolution of the entry-level smartphones in terms of AI deployment capabilities, we compare the Samsung A50 to the OnePlus Nord 2 5G. 
As a reference metric to measure the performance of each mobile device in the scenario of the deep learning models deployment, we provide the AI Score~\cite{Ignatov2018AIBR}. 
This AI score is computed over a set of experiments (\ie{ category recognition, semantic segmentation or image enhancement}), where inference precision and hardware-software behaviour are observed and quantified into a score metric depending on the importance of each of the factors.

\begin{figure}[!h]
    \centering
    \setlength{\tabcolsep}{2.0pt}
    \begin{tabular}{ccc}
    \includegraphics[trim=0 90 0 115,clip,width=0.325\linewidth]{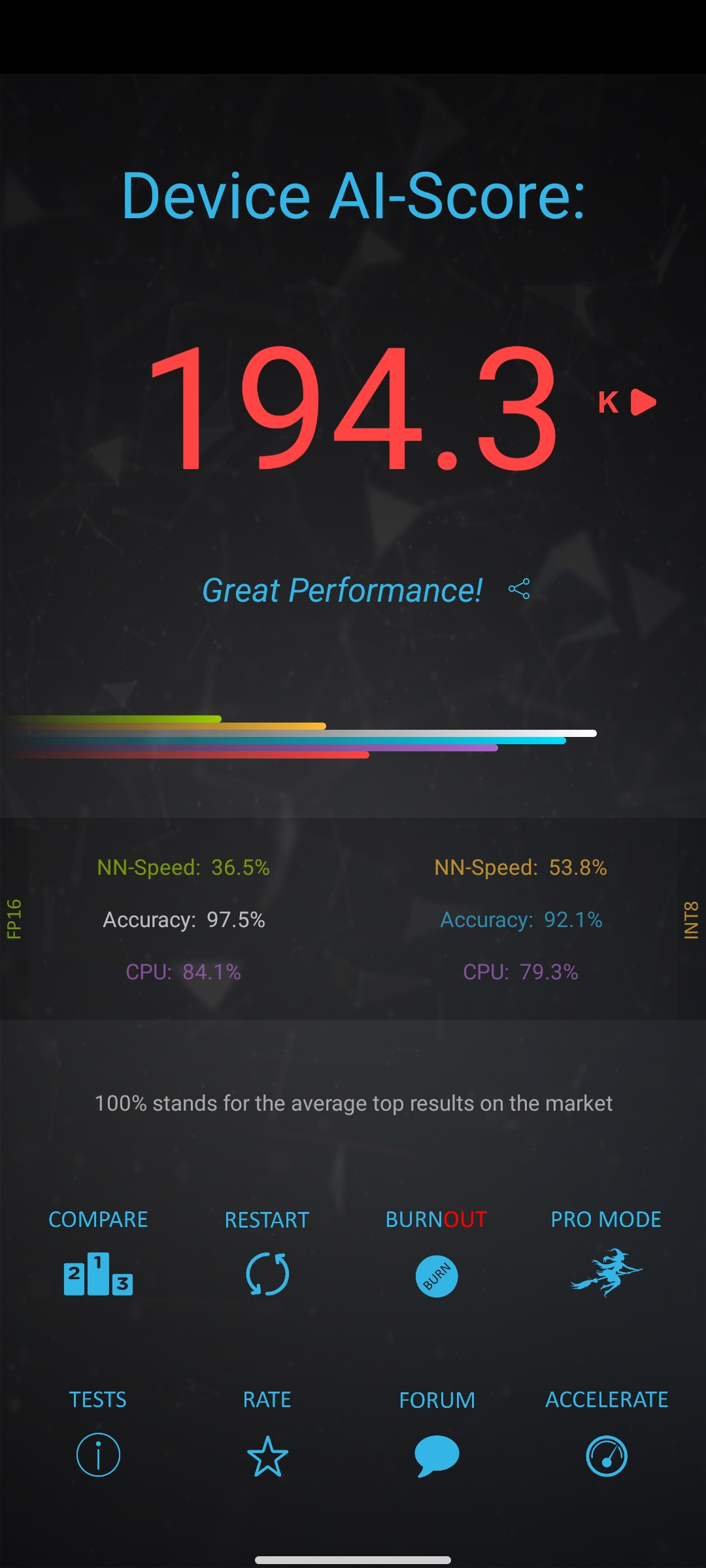} &
    \includegraphics[trim=0 90 0 115,clip,width=0.325\linewidth]{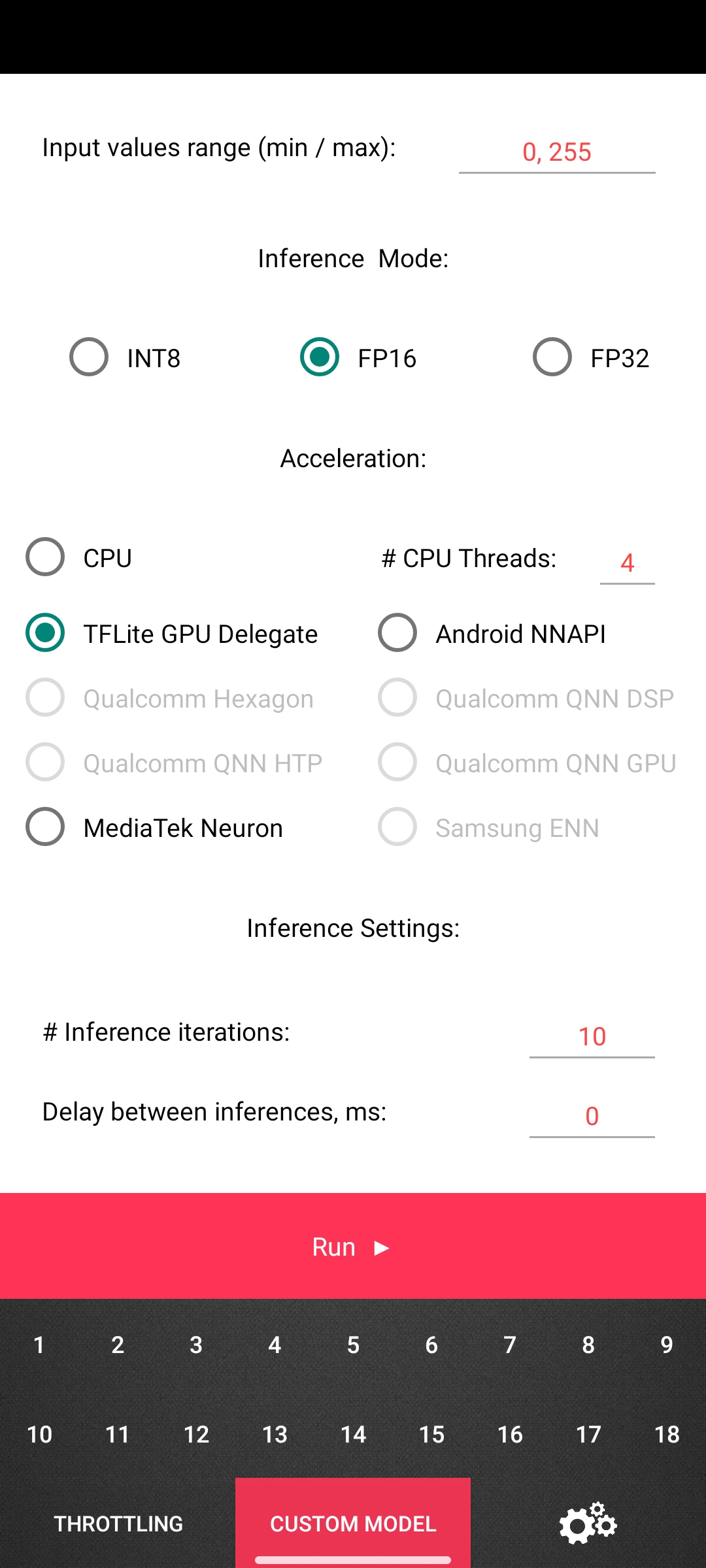} &
    \includegraphics[trim=0 90 0 115,clip,width=0.325\linewidth]{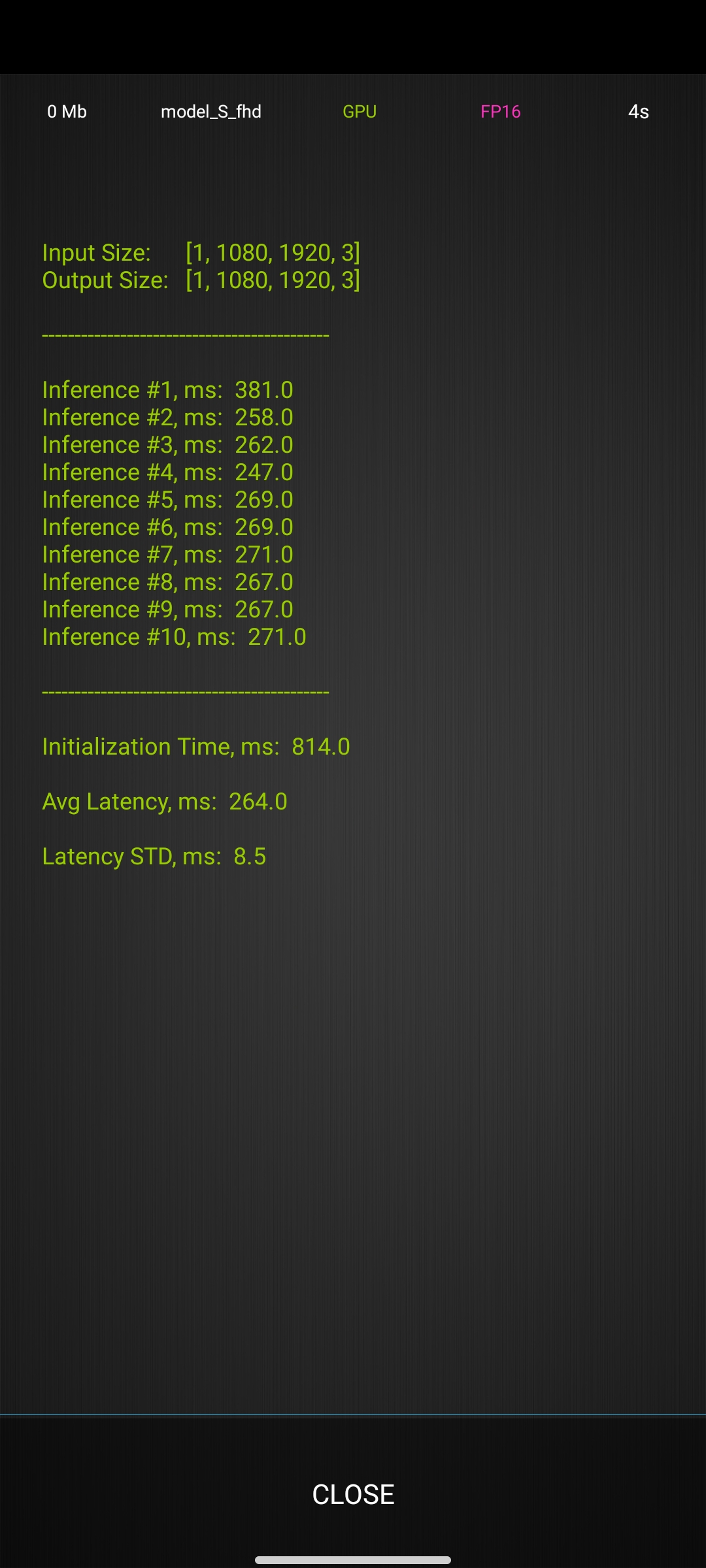} \tabularnewline
    \end{tabular}
    \caption{AI Benchmark~\cite{Ignatov2018AIBR} model performance evaluation.}
    \label{fig:aibench}
\end{figure}

All the tests are described by the authors in \cite{Ignatov2018AIBR}, and are available for the Android smartphones through the application \emph{AI Benchmark} \cite{Ignatov2018AIBR}, which also provides support for deep learning models deployment on different hardware options (CPU vs. GPU), and with various precision types (\ie{} FP16, FP32).
Models are tested on CPU and GPU, and FP16 precision after default Tensorflow-lite conversion and optimization, using as input a tensor image of different resolutions. The FLOPs are calculated at each resolution. We show in Figure~\ref{fig:aibench} the evaluation process using \emph{AI Benchmark app}~\cite{Ignatov2018AIBR}.

DISCNet~\cite{Feng_2021_CVPR_Discnet} cannot be directly converted to the required TFlite format due to unsupported operations. For this reason, we use DISCNet ``baseline (a)"~\cite{Feng_2021_CVPR_Discnet} to evaluate the model, this provides a fair lower-bound of the complete method's performance. Note that for an input of resolution $800\times800$, the authors report 364 GFLOPs~\cite{Feng_2021_CVPR_Discnet}, meanwhile in our re-implementation we obtain 442 GFLOPs\footnote{FLOPs measured using keras-flops library}. However, for a fair comparison, we compare with a different possible implementation with fewer operations.
SRMDNF~\cite{udc-ir-cvpr21} and SFTMD~\cite{udc-ir-cvpr21} cannot be directly converted to the requested standard format~\cite{Ignatov2018AIBR}, therefore, we use a canonical UNet model with the same number of FLOPs and memory requirements to approximate their performance. 

As we show in Table~\ref{tab:ablation}, our proposed model \emph{LUDCNet} can process Full-HD images in real-time in commercial mobile devices GPUs, being $\times5$ faster than DISCNet~\cite{Feng_2021_CVPR_Discnet}. Our model can also process low-resolution images on CPU in real-time (under 1 second per image), and its performance only decays $0.04$ with respect to DISCNet~\cite{Feng_2021_CVPR_Discnet} in terms of SSIM~\cite{SSIMPaper}. 

\begin{figure}[!ht]
    \centering
    \setlength{\tabcolsep}{2.0pt}
    \begin{tabular}{ccc}
    \includegraphics[width=0.32\linewidth]{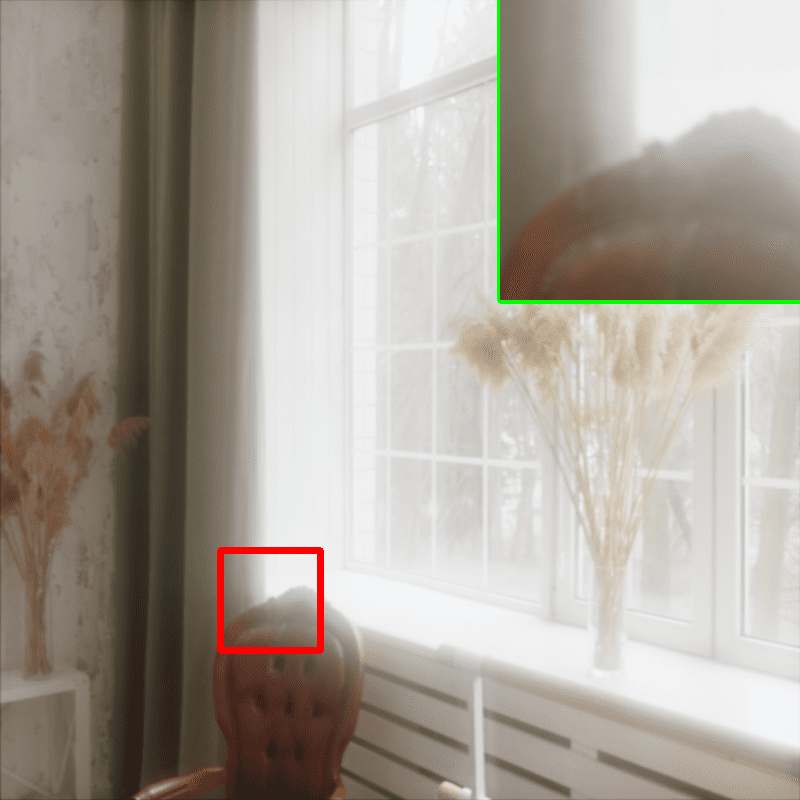} &
    \includegraphics[width=0.32\linewidth]{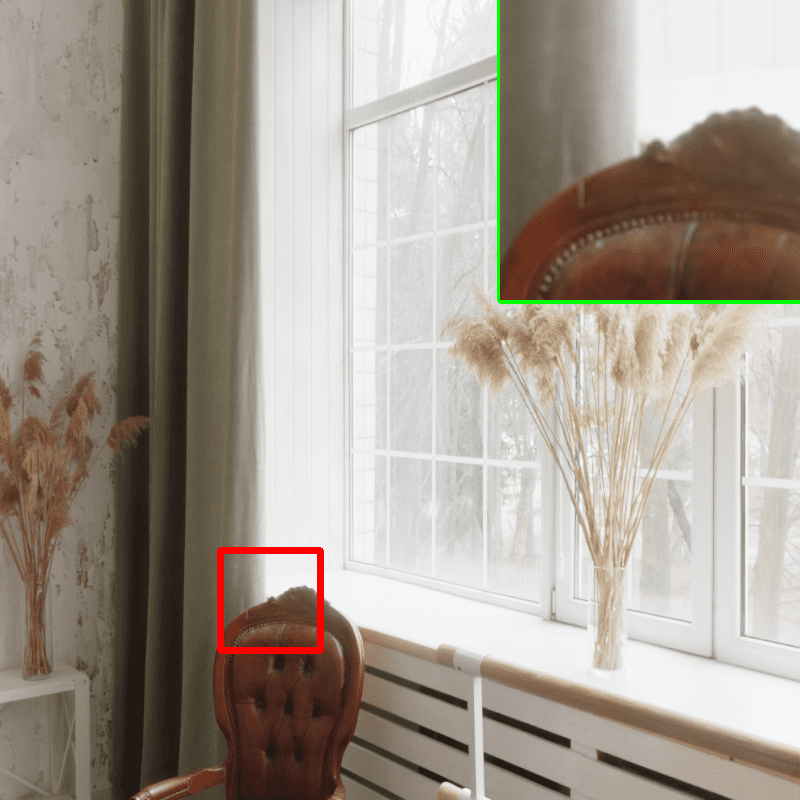} &
    \includegraphics[width=0.32\linewidth]{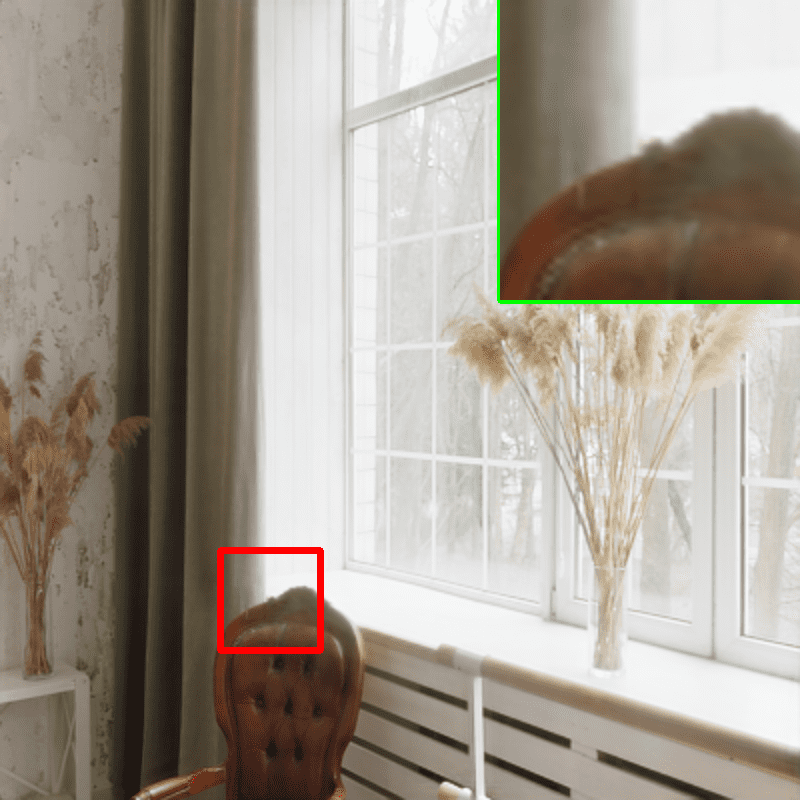} 
    \tabularnewline
    \includegraphics[width=0.32\linewidth]{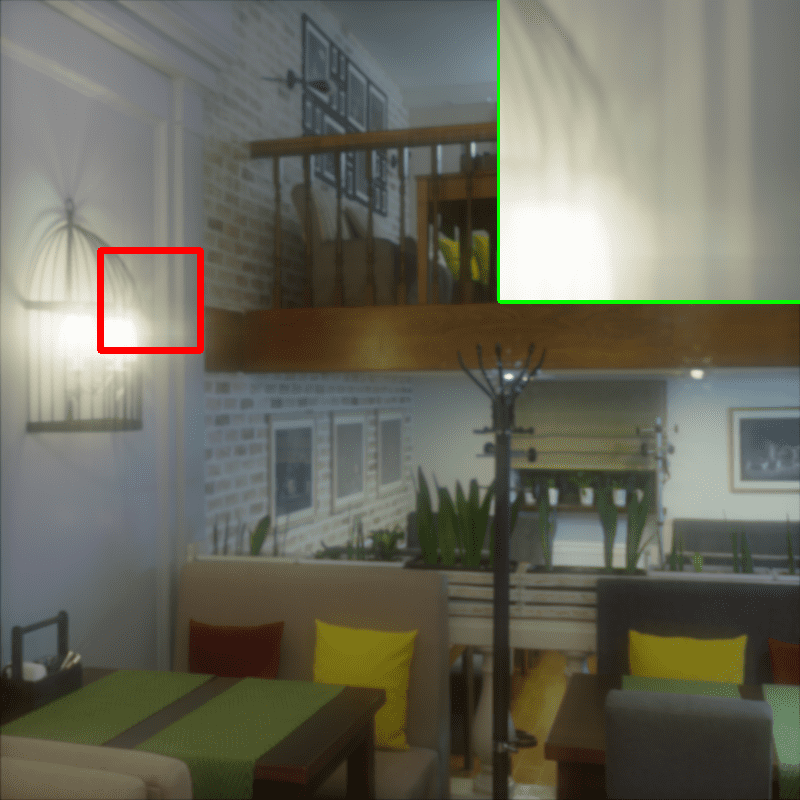} &
    \includegraphics[width=0.32\linewidth]{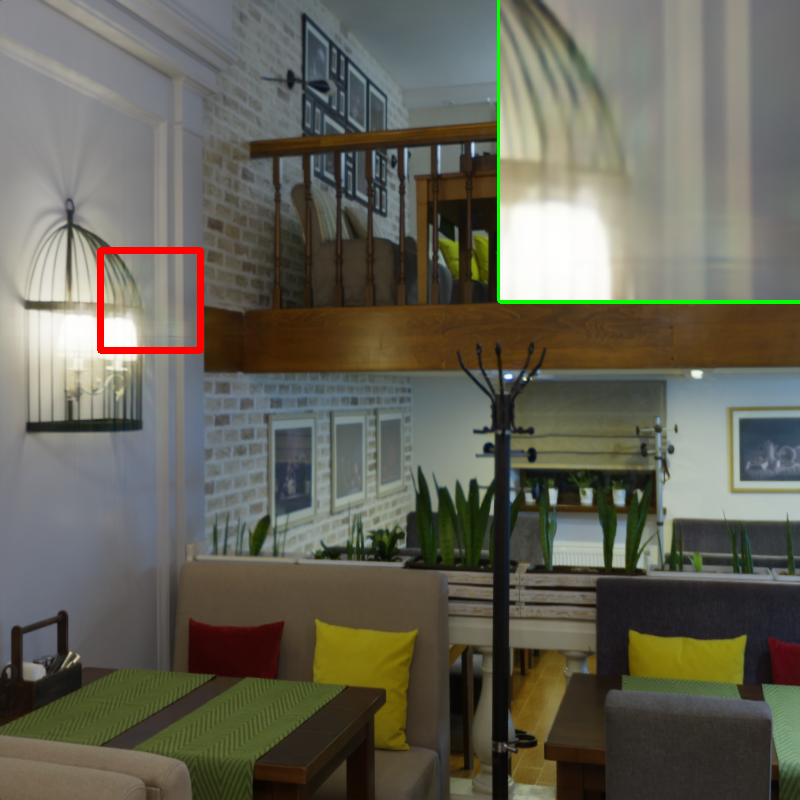} &
    \includegraphics[width=0.32\linewidth]{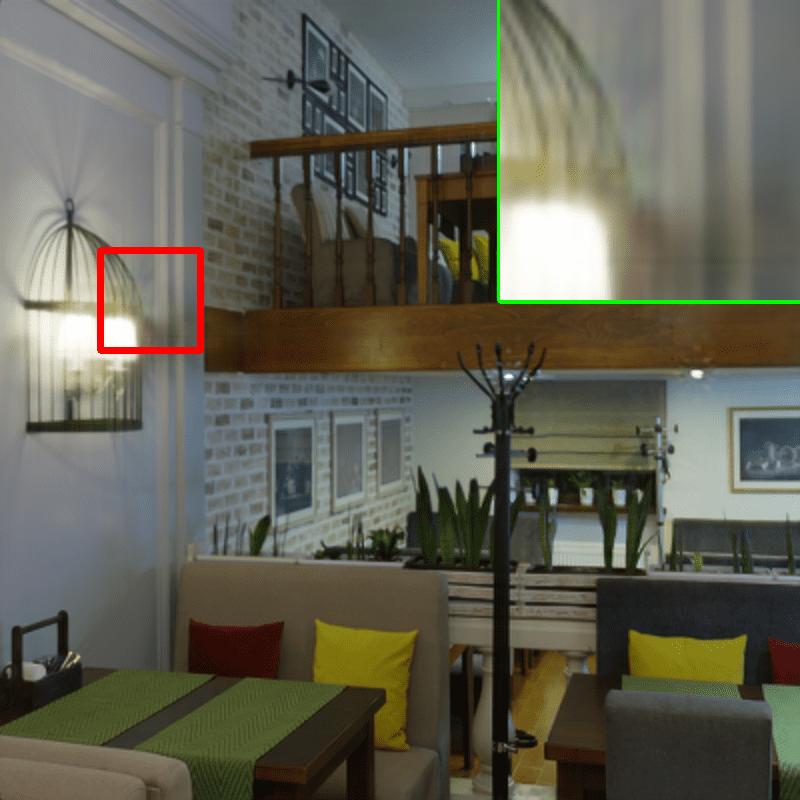}
    \tabularnewline
    \includegraphics[width=0.32\linewidth]{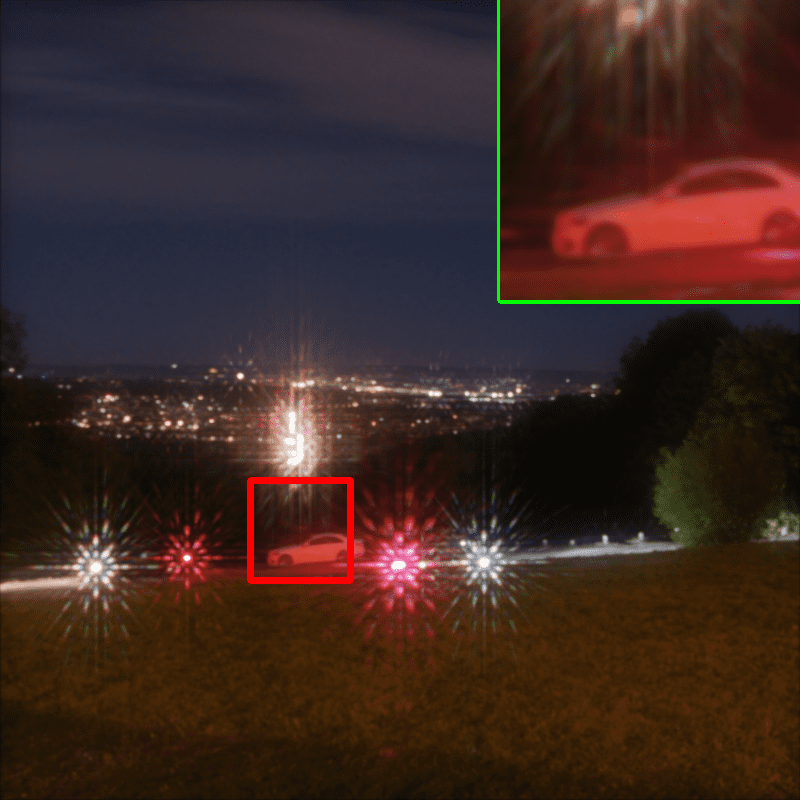} &
    \includegraphics[width=0.32\linewidth]{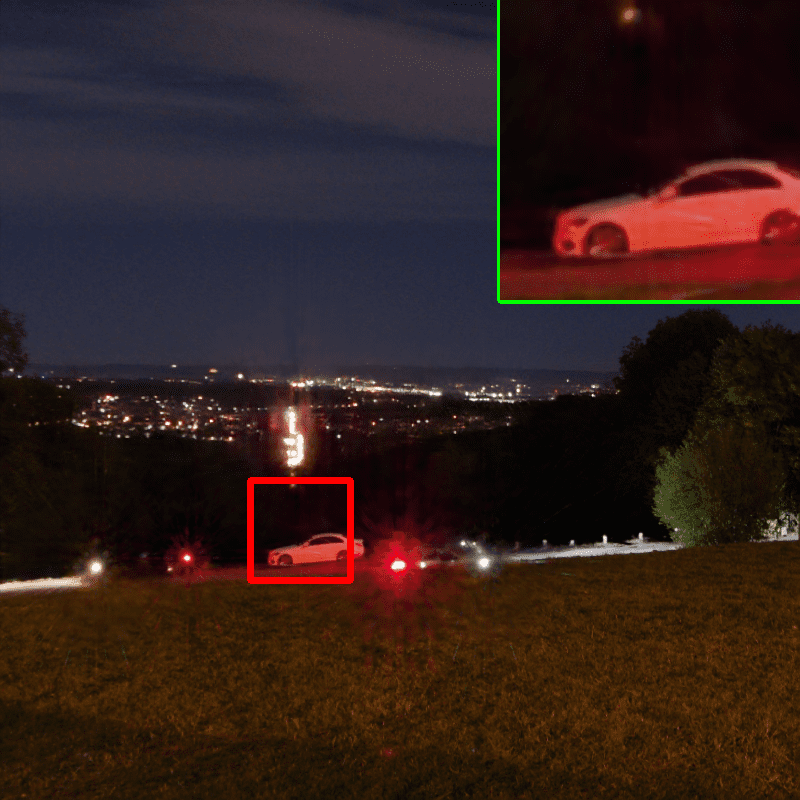} &
    \includegraphics[width=0.32\linewidth]{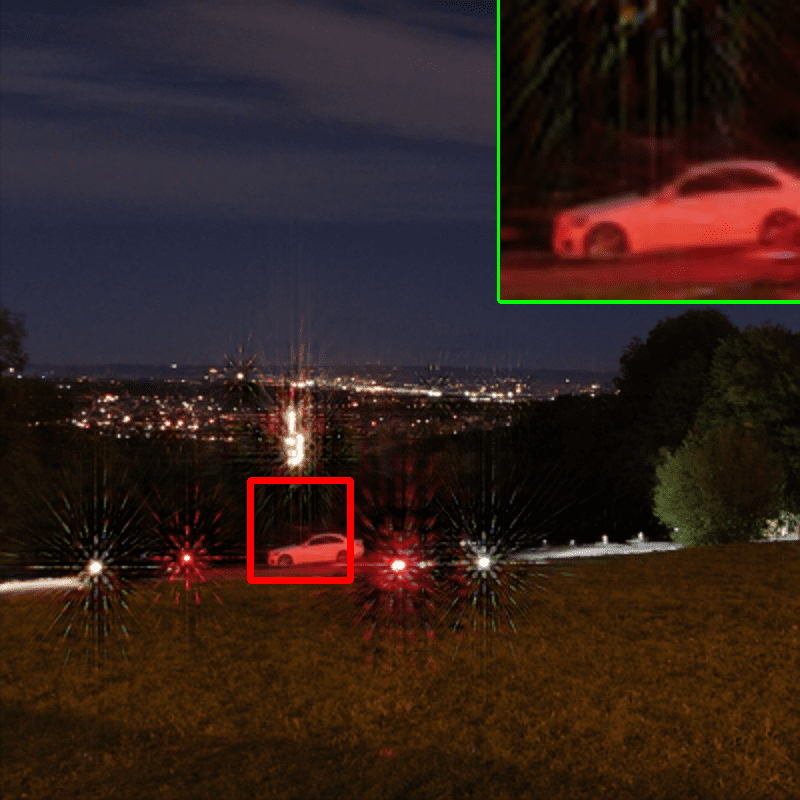}
    \tabularnewline
    \includegraphics[width=0.32\linewidth]{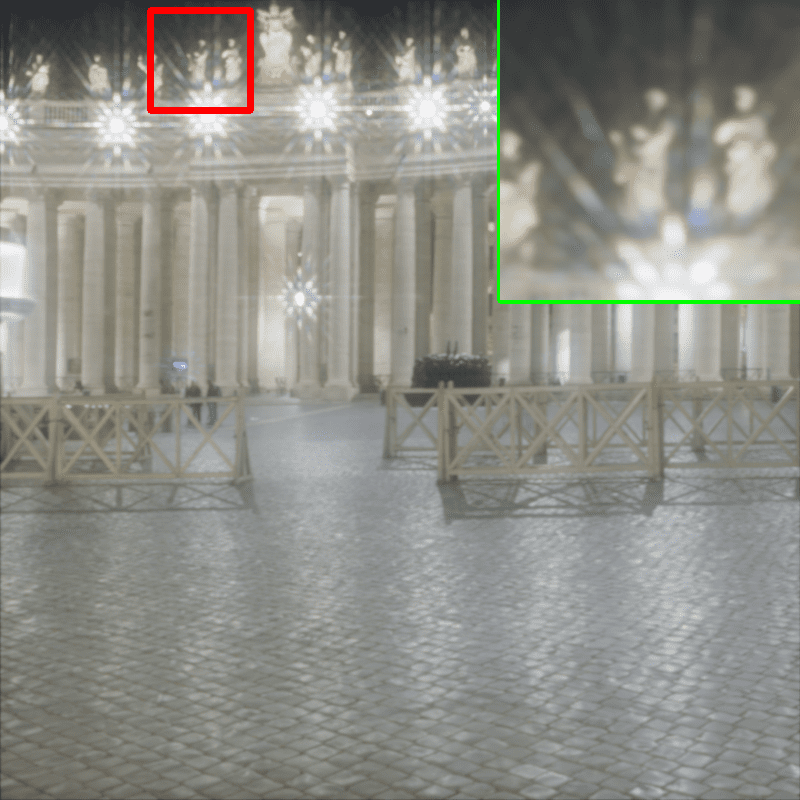} &
    \includegraphics[width=0.32\linewidth]{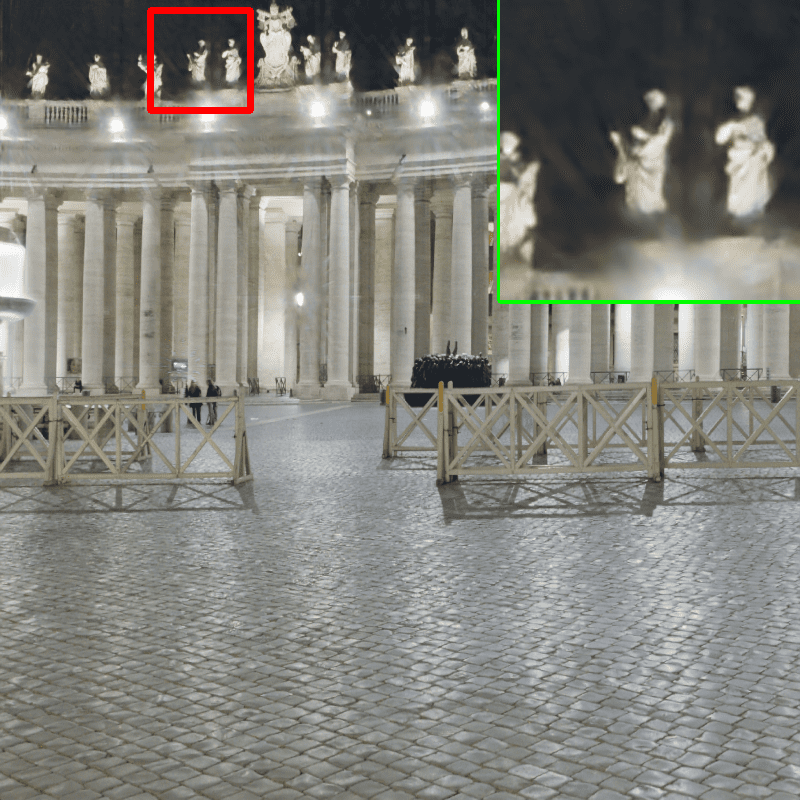} &
    \includegraphics[width=0.32\linewidth]{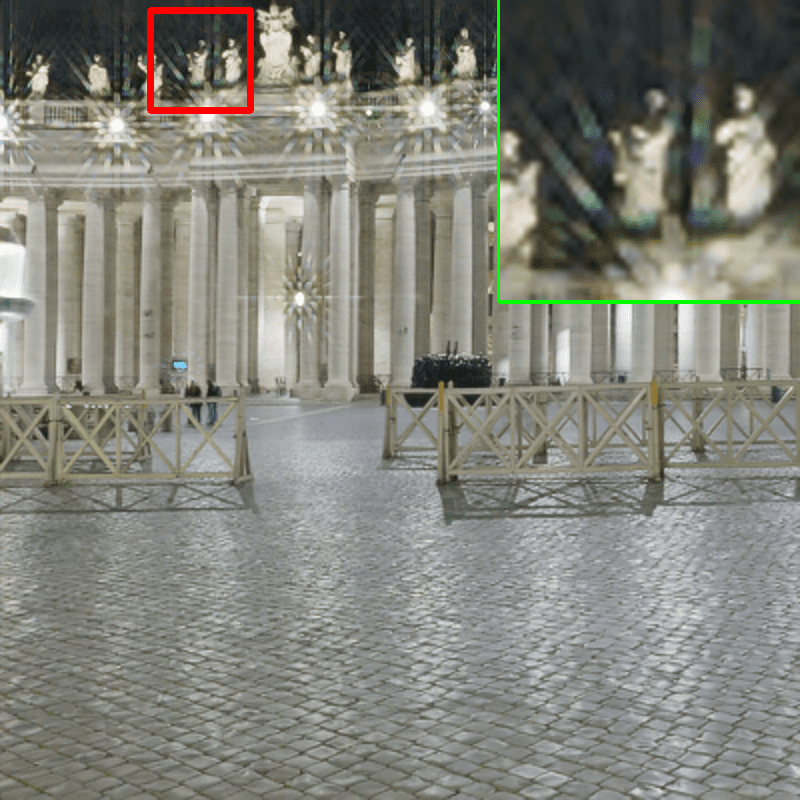}
    \tabularnewline
    Input & DRM-UDCNet & LUDCNet  \tabularnewline
    \end{tabular}
    \caption{Qualitative samples from the SYNTH test benchmark \cite{Feng_2021_CVPR_Discnet}. Our models can improve substantially the quality of the input UDC degraded images in indoor and outdoor, natural or artificial illumination, and day-night scenarios. As we explain in Section~\ref{sec:eff}, our lightweight model \emph{LUDCNet} can process Full-HD images in \textbf{real-time}.}
    \label{fig:test-samples}
\end{figure}

In Figure~\ref{fig:test-samples} we provide qualitative samples in challenging scenarios, our model is able to improve notably the perceptual quality of the UDC degraded images in real-time. Note that in the perception-distortion tradeoff~\cite{blau2018perception}, we focus on perceptual metrics~\cite{gu2022ntire, conde2022conformer}, as we aim for pleasant results for users.

\begin{table}[!ht]
    \centering
    \caption{Description of the selected commercial smartphone devices.}
    \label{tab:phones}
    \resizebox{\linewidth}{!}{
    \begin{tabular}{l|c|c|c|c|c|c}
        \hline\noalign{\smallskip}
        Phone Model & Launch & Chipset & CPU & GPU & RAM (GB) & AI Score~$\uparrow$ \\
        \hline
        (\# 1) Samsung A50 & 03/2019 & Exynos 9610 & 8 cores & Mali-G71 GP3 & 4 & 45.4  \\
        (\# 2) OnePlus Nord 2 5G & 07/2021 & MediaTek Dimensity 1200 & 8 cores & Mali-G77 MC9 & 8 & 194.3 \\
        (\# 3) OnePlus 8 Pro & 04/2020 & Qualcomm Snapdragon 865 5G & 8 cores & Adreno 650 & 12 & 137.0 \\
        \hline
    \end{tabular}
    }
\end{table}

\begin{table}[!h]
    \centering
    \caption{Efficiency benchmark for UDC Image Restoration on different commercial smartphones. We show the performance of different SOTA methods in terms of runtime at different image resolutions, device architectures and running scenarios (CPU, GPU). Runtimes are the average of at least 5 iterations. SRMDNF and SFTMD~\cite{udc-ir-cvpr21} failed the test ($\xmark$) due to exceed the memory limit. Our method is the only one that can perform Full-HD real-time processing, while we achieving high perceptual quality results.
    SSIM perceptual metric results are reported for the SYNTH dataset~\cite{Feng_2021_CVPR_Discnet}.}
    \label{tab:ablation}
    \resizebox{\linewidth}{!}{
    \begin{tabular}{l|c|c|c||c|c||c|c||c|c}
        \hline\noalign{\smallskip}
        & & &  &\multicolumn{6}{c}{Runtime (s)} \\
        Method & FLOPs~$\downarrow$ & Resolution &  SSIM~$\uparrow$ &\multicolumn{2}{c||}{Phone \#1} & \multicolumn{2}{c||}{Phone \#2} & \multicolumn{2}{c}{Phone \#3} \\
         name & (G) & (px) &  & CPU & GPU & CPU & GPU & CPU & GPU \\
         \hline
         SRMDNF~\cite{udc-ir-cvpr21} & 951 & 800$\times$800 & 0.965 & \xmark & \xmark & \xmark & \xmark & \xmark & \xmark \\
         SFTMD~\cite{udc-ir-cvpr21} & 2460 & 800$\times$800 & 0.986 & \xmark & \xmark & \xmark & \xmark & \xmark & \xmark \\
         \hline
         DISCNet~(a)~\cite{Feng_2021_CVPR_Discnet} & 442 & 800$\times$800 & 0.974  & 52.4 & 4.0 & 13.5 & 0.63 & 15.6 & 1.4\\
         DISCNet~(a)*~\cite{Feng_2021_CVPR_Discnet} & 300 & 800$\times$800 & \multirow{3}{*}{0.974}   & 13.4 & 2.3 & 4.4 & 0.37 & 5.5 & 0.9\\
         DISCNet~(a)~\cite{Feng_2021_CVPR_Discnet} & 256 & 256$\times$256 & & 1.1 & 0.41 & 0.3 & 0.07 & 0.6 & 0.14 \\
         \rowcolor{Gray} DISCNet~(a)~\cite{Feng_2021_CVPR_Discnet} & 1434 & 1920$\times$1080 &  & 1055 & 13.7  & 261.5 & 2.0 & 61.1 & 5.3 \\
         \hline
         \rowcolor{Gray} LUDCNet & \text100 & 1920$\times$1080 & \multirow{3}{*}{0.930} & 9.1 & \textbf{1.6} & 3.1 & \textbf{0.26} & 3.8 & \textbf{0.5} \\
         LUDCNet & 30 & 800$\times$800 &  & 2.6 & 0.5 & 0.9 & 0.095 & 1.2 & 0.19 \\
         LUDCNet & 3 & 256$\times$256 &  & 0.3 & 0.05 & 0.1 & 0.023 & 0.1 & 0.02 \\
         \hline
    \end{tabular}}
\end{table}

\begin{figure}[!h]
    \centering
    \includegraphics[width=0.62\linewidth]{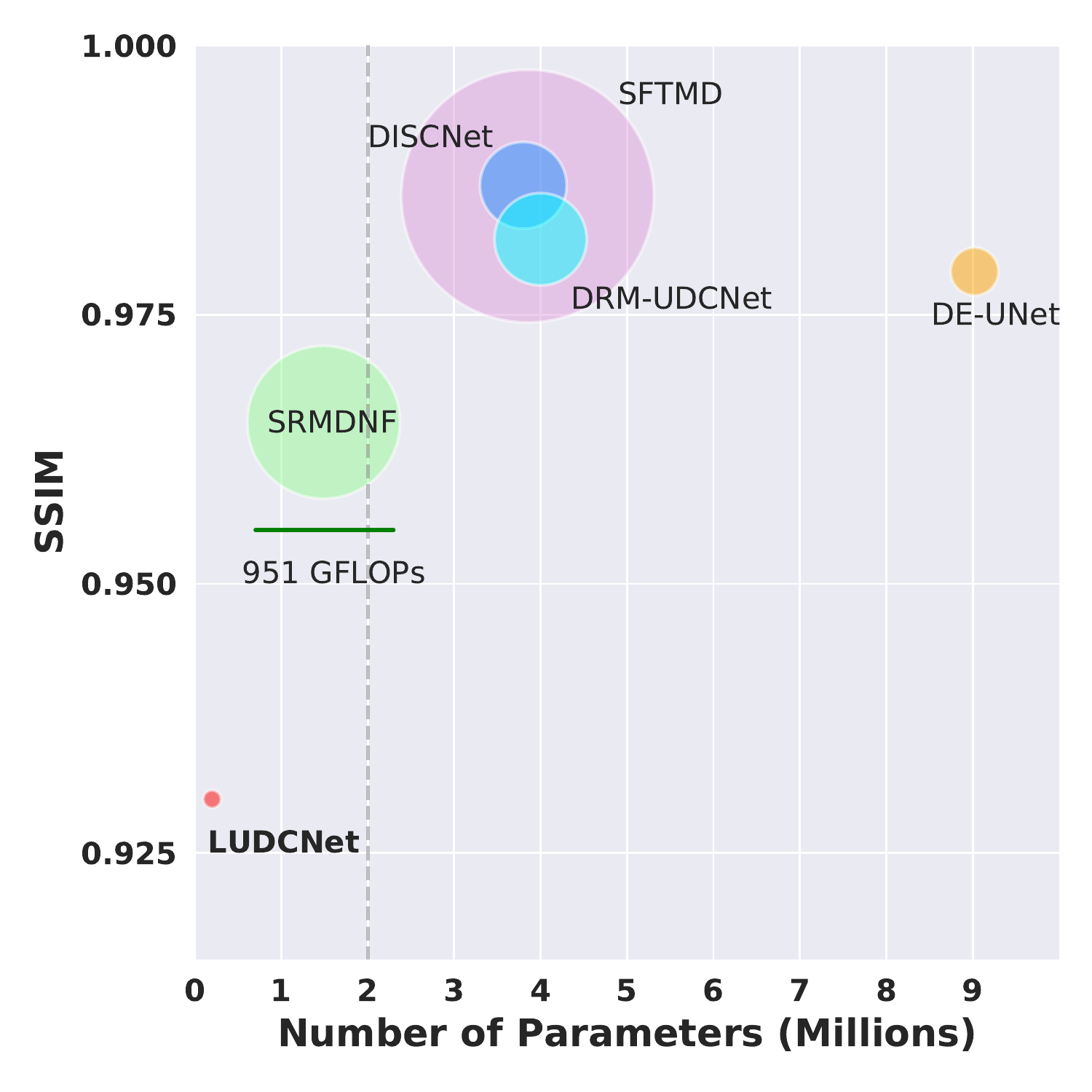}
    \caption{Comparison of parameters, FLOPs and performance of state-of-the art methods for UDC Image Restoration. Our method achieves high perceptual quality results while being notable smaller and computationally ``cheap".}
    \label{fig:flops_params}
\end{figure}


\vspace{1mm}
\noindent\textbf{Perceptual Image Enhancement for Smartphone Applications}\\ Besides the proposed model, we also developed a considerably more powerful architecture called LPIENet \cite{conde2022perceptual}, and a extended efficiency benchmark~\cite{conde2022perceptual}. Our new model can process 2K resolution images under 1 second in mid-level commercial smartphones, and tackles: denoising, deblurring and UDC image restoration.

\section{Conclusions}

First, we develop DRM-UDCNet, a U-Net based model to solve this problem and achieve the most competitive performance on well-known benchmarks. Next, we focus on compacting such model, aiming to be the first work to tackle real-time UDC image restoration in the smartphone deployment scenario. As a result, our model LUDCNet uses $\times4$ less operations than other state-of-the-art methods. 
Efficiency analysis shows that our compact model is $\times5$ faster than DISCNet. We set up a new benchmark to show our method's potential, using three commercial smartphones showcasing different hardware architectures and capabilities.
We show that our model does not depend on specific hardware, achieving real-time performance on current mobile phone CPUs at lower resolutions. 
To the best of our knowledge, we are the first work to approach and analyze the UDC restoration problem from the efficiency and production point of view.

\vspace{2mm}
\noindent\textbf{Acknowledgments}
This work was partly supported by The Alexander von Humboldt Foundation (AvH).

\clearpage



%
%
\bibliographystyle{splncs04}
\bibliography{egbib}

\end{document}